\documentclass[%
aip,
 amsmath,amssymb,
 reprint,%
]{revtex4-1}

\usepackage{graphicx}
\usepackage{dcolumn}
\usepackage{bm}

\usepackage[utf8]{inputenc}
\usepackage[T1]{fontenc}
\usepackage{mathptmx}
\usepackage{etoolbox}

\makeatletter
\def\@email#1#2{%
 \endgroup
 \patchcmd{\titleblock@produce}
  {\frontmatter@RRAPformat}
  {\frontmatter@RRAPformat{\produce@RRAP{*#1\href{mailto:#2}{#2}}}\frontmatter@RRAPformat}
  {}{}
}%
\makeatother
\begin{document}

\preprint{AIP/123-QED}

\title[]{Star-shaped nanostructure-based multifunctional chiral metasurface for near-infrared wavelengths}
\author{Md. Ehsanul Karim}
 \affiliation{ 
Department of Electrical and Electronic Engineering, Bangladesh University of Engineering and Technology, Dhaka- 1205, Bangladesh
}
 \affiliation{Department of Electrical and Electronic Engineering, BRAC University, Dhaka- 1212, Bangladesh} 
\author{Ahmed Zubair}%
 \email{ahmedzubair@eee.buet.ac.bd}
\affiliation{ 
Department of Electrical and Electronic Engineering, Bangladesh University of Engineering and Technology, Dhaka- 1205, Bangladesh
}%

\date{\today}

\begin{abstract}
Here, we reported, for the first time, a chiral metasurface with multifunctional capabilities tailored for the near-infrared (NIR) wavelength range, garnering tremendous interest in applications such as optical communication and chiral sensing. The structure comprised star and bar-shaped silver antennas immersed in $\mathrm{SiO_2}$. In computational simulations based on the finite-difference time-domain method, the proposed architecture yielded a maximum circular dichroism of 0.7 at 971.5 nm wavelength and a maximum asymmetric transmission under linearly polarized excitation of 0.92 at 1020 nm wavelength. Moreover, this multifunctional structure manifested near-perfect cross-polarization conversions for both linear and circular polarizations, as well as linear to circular and elliptical polarization conversion effects at different bands within the NIR spectral region. The enhanced light-matter interaction of the proposed structure was utilized to detect and differentiate two enantiomers of 1,2-propanediol, yielding excellent chirality sensing capability. As a multifunctional compact chiral platform, the proposed structure is expected to find direct applications in integrated and miniaturized multi-channel optical communication platforms, materials and biomolecules sensing in attomolar concentrations, and the pharmaceutical industry.
\end{abstract}

\maketitle

%

\section{Introduction}
Chirality refers to the non-superimposable feature of a molecule, ion, crystal, or device on its mirror image \cite{papakostas2003optical}. Since its discovery, this property of natural materials, namely DNA, proteins, amino acids, enzymes, and crystals \cite{utembe2019chirality}, has been utilized in different applications like drug and fragrance discovery \cite{brandt2017added} and estimation of protein secondary structure \cite{greenfield2006using}. However, the weak chiro-optical response of the natural materials means the detectable chiral activity can only be generated from such entities if the light wavelength is much smaller than the optical path length \cite{wang2016optical}.

Metasurfaces, with their unique ability to manipulate the amplitude, phase, and polarization of electromagnetic waves \cite{kim2022tunable}, have garnered significant interest from the photonics research community over the last decade in applications like cloaking \cite{wei2017ultrathin}, lensing \cite{ndao2020octave}, absorption \cite{Partha23NA}, light manipulation\,\cite{Sarker:24}, holography \cite{zheng2015metasurface}, and sensing\,\cite{Sarker24PCCP}. The chiral subset of these nanoscale structures, comprising artificial atoms lacking in-plane mirror symmetry \cite{de2015strong}, can enhance the weak chiro-optical response of the natural media. They have been used extensively for producing non-trivial optical phenomena like circular/linear asymmetric transmission \cite{shen2022chiral}, generation and manipulation of linear or circularly polarized light \cite{fei2020versatile,mun2019broadband}, orbital angular momentum generation \cite{lin2022chirality}, and polarization sensitive retro-reflection \cite{wang2021spin} in a compact configuration. With the ever-increasing requirement for compactness and integration in complex photonic platforms, there has been an increased effort from the scientific community to integrate several of these functionalities in the same meta-device. Several efforts for realizing chiral multifunctional metasurfaces manifesting a diverse range of functionalities have been reported in previous literature for mainly GHz and THz frequencies \cite{yin2023multi,habashi2023bi}.

Recently, with the improvement of nanofabrication techniques, chiral metasurfaces and metamaterials in general have experienced a shift towards higher frequencies \cite{cao2021infrared}. Chiral metasurfaces in the near-infrared (NIR), with their enhanced chiro-optical response, can be a valuable tool for chiral sensing, as many of the chiral molecules exhibit resonant behavior at this spectral band \cite{wang2016novel}. Another important avenue for applications of NIR chirality is multi-channel optical communication \cite{zhao2012twisted,cheng2015emergent}, where their ability to produce asymmetry in transmission and polarization manipulation can be used for noise cancellation \cite{sounas2017non}, source isolation \cite{dixon2021self}, signal processing \cite{kim2020giant}, information encryption \cite{jiang2023linear}, and waveplating \cite{hu2017all}. Consequently, several structures investigating NIR chirality can be found in recent reports \cite{ali2023dielectric,wu2022near,zong2022multiple,ouyang2018near,li2019strong,mao2019extrinsically}. Recently, Liu \textit{et al.} reported a NIR chiral meta-device with asymmetric transmission (AT) for both linear and circular polarizations \cite{liu2019dual}. However, the maximum AT reported in the work is relatively low compared to other state-of-the-art structures in this spectral band. Moreover, polarization manipulation, a distinctive and key feature of many chiral applications, was not realized with this device. To our knowledge, a NIR-range chiral metasurface with a wide range of functionalities has not been reported.

In this paper, we proposed a multifunctional chiral metasurface in the NIR regime. The structure's unit cell comprised star and bar-shaped silver (Ag) nanoantennas immersed in $\mathrm{SiO_2}$ dielectric. The final topology was achieved based on iterative tuning of structural parameters and antenna shape to maximize circular dichroism (CD), quantifying the circular handedness selective transmission of the structure. By changing the incidence wavelength and polarization, we demonstrated linear and circular asymmetric transmission, circular and linear cross-polarization switching, and linear to circular and elliptical polarization conversion capability of our proposed device in the near-IR regime. This marks the first report of a NIR chiral metastructure with multifunctionalities. Additionally, we investigated the electric field profiles of the structure to gain insight into the origin of the asymmetry in transmission for linear and circular polarization handedness. The enhanced chirality of the proposed metasurface was utilized as a tool for handedness differentiation of the two enantiomers of 1,2-propanediol. The ability of the device to produce distinctive optical signals for the two isomers, overcoming the inherently weak response of the standalone chiral molecules, suggested the suitability of the proposed device in chiral sensing applications. We expect our star-shaped multifunctional chiral metasurface to pave the way for integrated and miniaturized photonic platforms in NIR multichannel communication and chiral sensing systems.

\section{Methodology and structural design}
\label{sec:unit_cell}

\begin{figure*}
\includegraphics[width=5in]{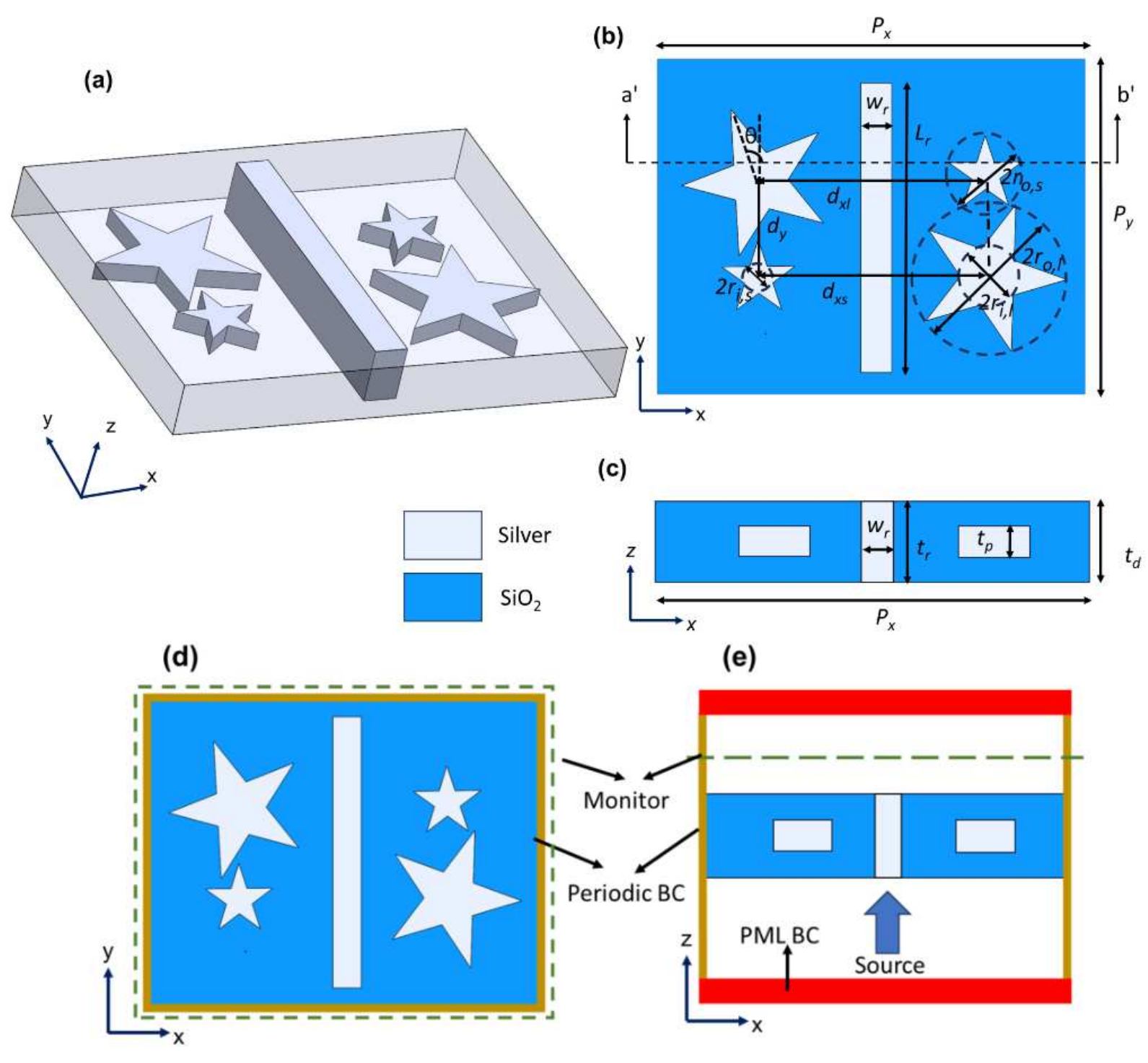}
\caption{(a) 3D schematic representation of the star-shaped antenna-based metasurface. The Ag rod and stars are immersed in the dielectric ($\mathrm{SiO_2}$), which has been made transparent here for easy visualization. (b) x-y and (c) x-z plane cross-sectional view of the star-shaped metasurface, with different dimensional parameters pointed out. The periodicity along x and y are denoted as $P_x$ and $P_y$, respectively. The dielectric thicknesses ($\mathrm{SiO_2}$) and stars are represented by $t_d$ and $t_p$ respectively. The dimensions of the Ag rod in the x, y, and z directions are $w_r$, $L_r$, and $t_d$, respectively. $d_{xs}$ and $d_{xl}$ are the separations between the centers of the small and large stars along the x direction. The dimensions of the stars are characterized by the radii of the two dotted circles through the inner and outer points. For the larger (smaller) stars (at the top-left(right) and bottom-right(left)), these parameters are $r_{i,l(s)}$ and $r_{o,l(s)}$, respectively. The larger stars are tilted with respect to their smaller counterpart by an angle $\theta$. Illustration of the simulation setup of the structure along a (d) x-y and (e) z-x cross-section with a plane wave illumination propagating along the z direction. }
\label{fig:schematic}
\end{figure*}

The 3D schematic representation of the unit cell of our designed star-shaped nanoantenna-based chiral metasurface is depicted in Fig. \ref{fig:schematic}(a). The structure consists of four Ag stars and an Ag bar embedded in a dielectric ($\mathrm{SiO_2}$). On each side of the bar along the x-direction, there is a pair of stars, one small and the other large (Fig. \ref{fig:schematic}(b)). This asymmetry allows us to produce different chiral characteristics in the final device. The optical characterization of the structure, including optimizing different geometrical topologies, was conducted using the Ansys Lumerical FDTD. In the simulation environment, periodic boundary conditions were deployed in the x and y directions, while perfectly matched layer (PML) boundary conditions were used along the third axis, as shown in Figs. \ref{fig:schematic}(d) and (e). A plane wave source propagating along the z-direction illuminates the structure. A 2D power monitor in the x-y plane measured the transmittances. We used the Pallik model for the optical properties of $\mathrm{SiO_2}$ \cite{palik1998handbook}. For Ag, the refractive index data from the Johnson and Christy model was utilized \cite{johnson1972optical}. 

\section{Results and discussion}

\subsection{Tuning the structural parameters and shape of antennas}
\label{sec:tuning}

\begin{figure*}
\includegraphics[width=18cm]{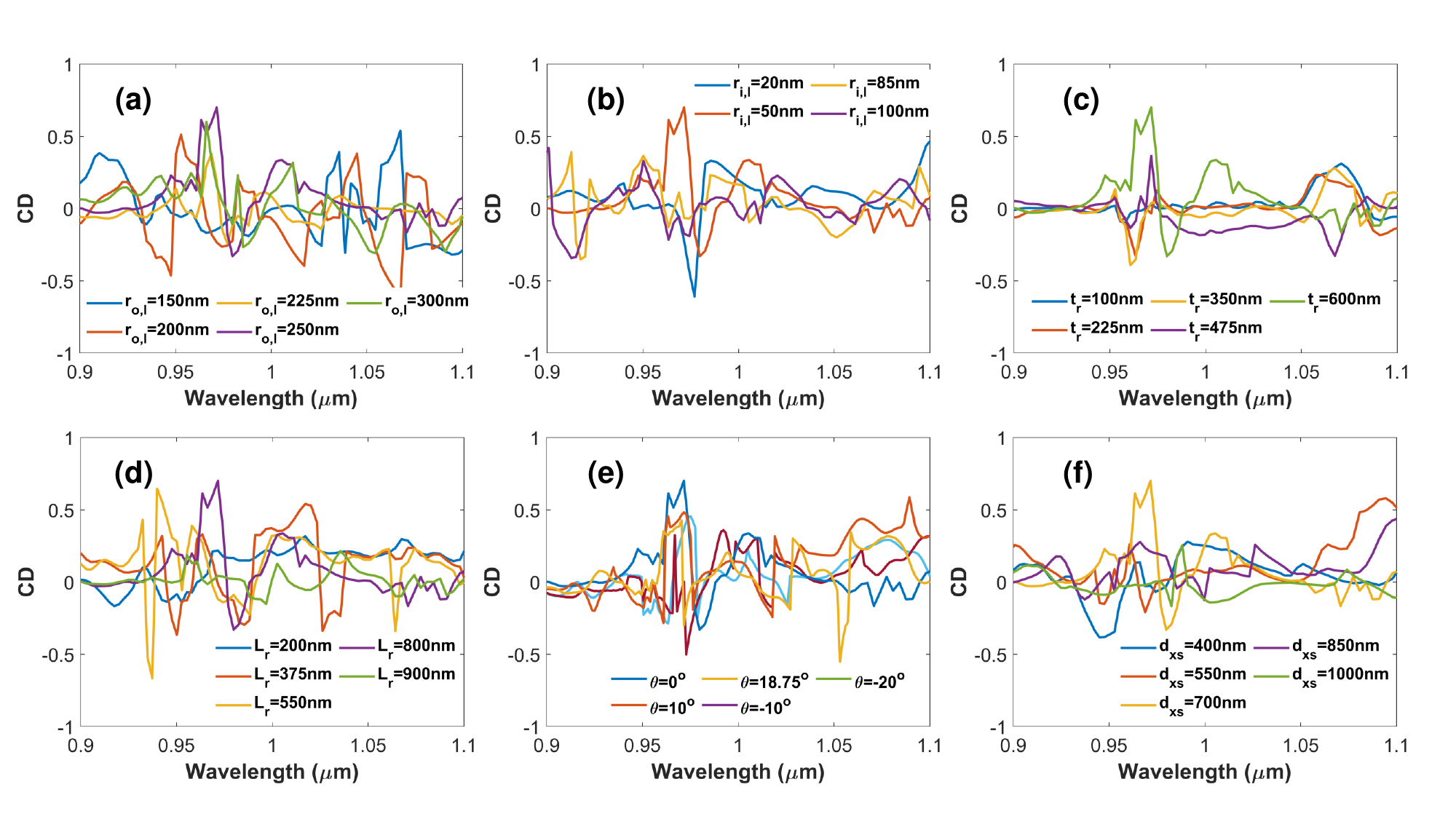}
\caption{CD spectra of the star-shaped antenna-based structure with different values of \textbf{(a)} $r_{o,l}$, \textbf{(b)} $r_{i,l}$, \textbf{(c)} $t_r$, \textbf{(d)} $L_r$, \textbf{(e)} $\theta$, and (f) $d_{xs}$. All the dimensional parameters were kept constant at their optimal values.}
\label{fig:sweep1}
\end{figure*}

\begin{figure*}
\centerline{\includegraphics[width=1.1\textwidth]{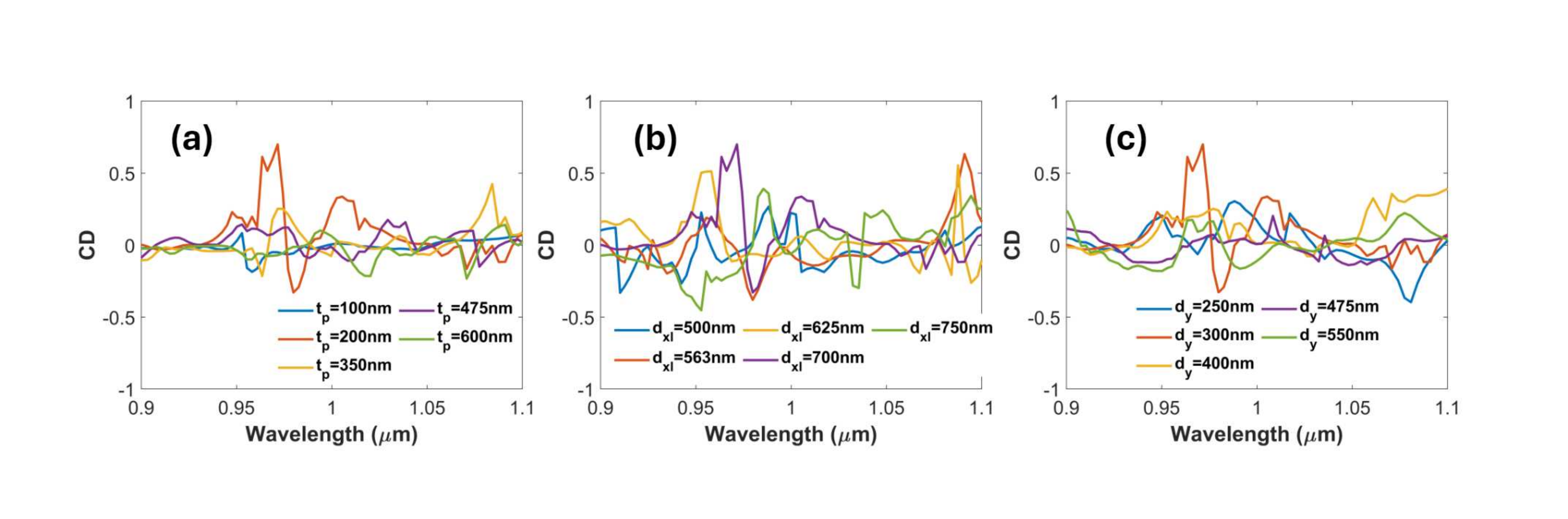}}
\caption{CD spectra of the star-shaped antenna-based structure with different values of \textbf{(a)} $t_{p}$, \textbf{(b)} $d_{xl}$, and \textbf{(c)} $d_y$. All the dimensional parameters were kept constant at their optimal values.}
\label{fig:sweep2}
\end{figure*}

\begin{figure*}
\centerline{\includegraphics[width=13cm]{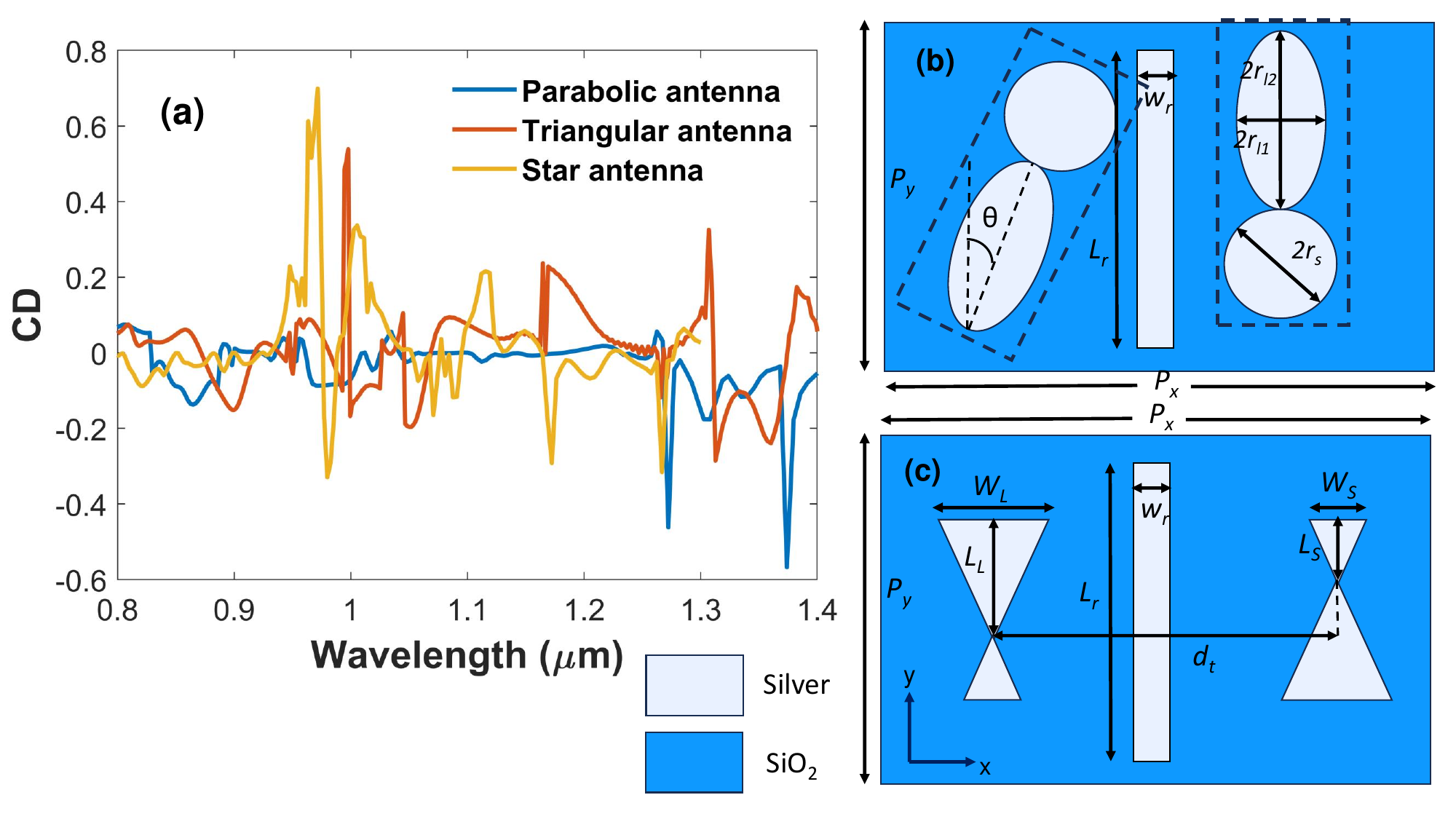}}
\caption{Effect of the nanoantenna shape on CD spectra of the metasurface. \textbf{(a)} CD spectra for parabolic, triangular, and star-shaped antenna. All the structures were optimized individually to maximize the corresponding CD value at the NIR frequencies. Section \ref{sec:unit_cell} showcases the final star-shaped structure and its parameters. x-y plane cross-section of the optimized metasurface with \textbf{(b)} parabolic and \textbf{(c)} triangular nanoantenna. For the structure in (b), the two dotted boxes represent the antennas' positions. The antenna pair on the left is tilted by an angle of $\theta$ to the ones on the right. The dimensions of the two axes of the two parabolic antennas are 2$r_{l2}$ and 2$r_{l2}$, while the radius of the circular antenna is $r_s$. The dimension of larger (smaller) triangles in (c) is represented by a height of $L_{L(S)}$ and a width of $W_{L(S)}$. $P_x$ and $P_y$ mark both structures' periodicities along the x and y directions. The height and width of the metal bar are marked by $L_r$ and $w_r$, respectively.}
\label{fig:shape_sweep}
\end{figure*}

We iteratively tuned different structural parameters of the star-shaped antenna-based metasurface structure to maximize circular dichroism (CD), defined in Eq. \ref{eq:CD}, in the NIR regime. The variation of the CD spectra with several parameters has been shown in Figs. \ref{fig:sweep1} and \ref{fig:sweep2}. Although we used the spectral region from 600 nm to 1.3 $\mu$m during our optimization process, the results shown here have a truncated spectral region for easy visualization. Based on these results, the structural parameters for the final unit cell are $P_x$=1.26 $\mu$m, $P_y$=950 nm, $t_d$=600 nm, $t_p$=200 nm, $w_r$=100 nm, $L_r$=800 nm, $r_{i,l}$=50 nm, $r_{o,l}$=250 nm, $r_{i,s}$=30 nm, $r_{o,l}$=150 nm, $d_{xs}$=$d_{xl}$=700 nm, $d_y$=300 nm, and $\theta$=18.75$^\circ$. 

Moreover, we experimented with two other geometrical topologies for the antennas, namely parabolic ((Fig. \ref{fig:shape_sweep}(b))) and triangular (Fig. \ref{fig:shape_sweep}(c)), using the same basic structure. We individually optimized both structures to maximize CD. The structural parameters for the optimized architecture with the parabolic-shaped antenna in Fig. \ref{fig:shape_sweep}(b) are $P_x$=1.26 $\mu$m, $P_y$=950 nm, $w_r$=100 nm, $L_r$=800 nm, $r_{l1}$=100 nm, $r_{l2}$=200 nm, $r_{s}$=160 nm, $\theta$=15$^\circ$, thicknesses of the triangular antenna, metal bar, and dielectric layer are 200, 600, and 600 nm, respectively. The positions of the antennas are defined in terms of the two dotted boxes encircling the pairs. Their centers are 500 and 50 nm apart along the x and y-directions, respectively. In the case of the tuned architecture with triangular-shaped antennas, different parameters are $P_x$=1.26 $\mu$m, $P_y$=950 nm, $w_r$=100 nm, $L_r$=800 nm, $W_L$=260 nm, $L_L$=300 nm, $W_S$=150 nm, $L_S$=150 nm, $d_t$=800 nm, thicknesses of the triangular antenna, metal bar, and dielectric layer are 200 nm, 225 nm, and 600 nm respectively. The comparison of the CD spectra for all three optimized structures (star, parabolic, and triangular) in Fig. \ref{fig:shape_sweep}(a) reveals the superior performance of the star-shaped antenna-based architecture. Therefore, our final design chose the tuned star-shaped architecture. Other optical performance parameters of this structure will be investigated later.

\subsection{Asymmetric transmission for linear polarized light}
We used the Jones matrix formulation to quantify the transmission characteristics of the star-shaped antenna-based structure under linearly polarized light illumination. It mathematically connects the input electrical fields ($E_x^{i}$, $E_y^{i}$) with their transmitted counterparts ($E_x^{t}$, $E_y^{t}$) through a 2$\times$2 matrix \cite{jones1947new}, 

\begin{equation}
\label{eq:jones_lin}
   \left[ \begin{array}{c}
         E_x^{t}   \\
        E_y^{t}  
    \end{array}\right] =
    \left[ \begin{array}{cc}
        t_{xx} & t_{xy} \\
         t_{yx}& t{yy}
    \end{array}
    \right]
    \left[ \begin{array}{c}
         E_x^{i}   \\
        E_y^{i}  
    \end{array}\right]= T_{lin} \left[ \begin{array}{c}
         E_x^{i}   \\
        E_y^{i}  
    \end{array}\right]
\end{equation}

The entries of the matrix $T_{lin}$ in Eq. \ref{eq:jones_lin} are called the co-polarization ($t_{xx}$, $t_{yy}$) and cross-polarization ($t_{xy}$, $t_{yx}$) transmission coefficients and are dispersive and complex in general. Here, the subscripts \textit{x} and \textit{y} represent electric field components polarized along the x and y directions, respectively. For instance, $t_{xy}$ represents the transmission co-efficient of x polarized light when a y polarized light is incident on the structure. 

\begin{figure*}
\centerline{\includegraphics[width=15cm]{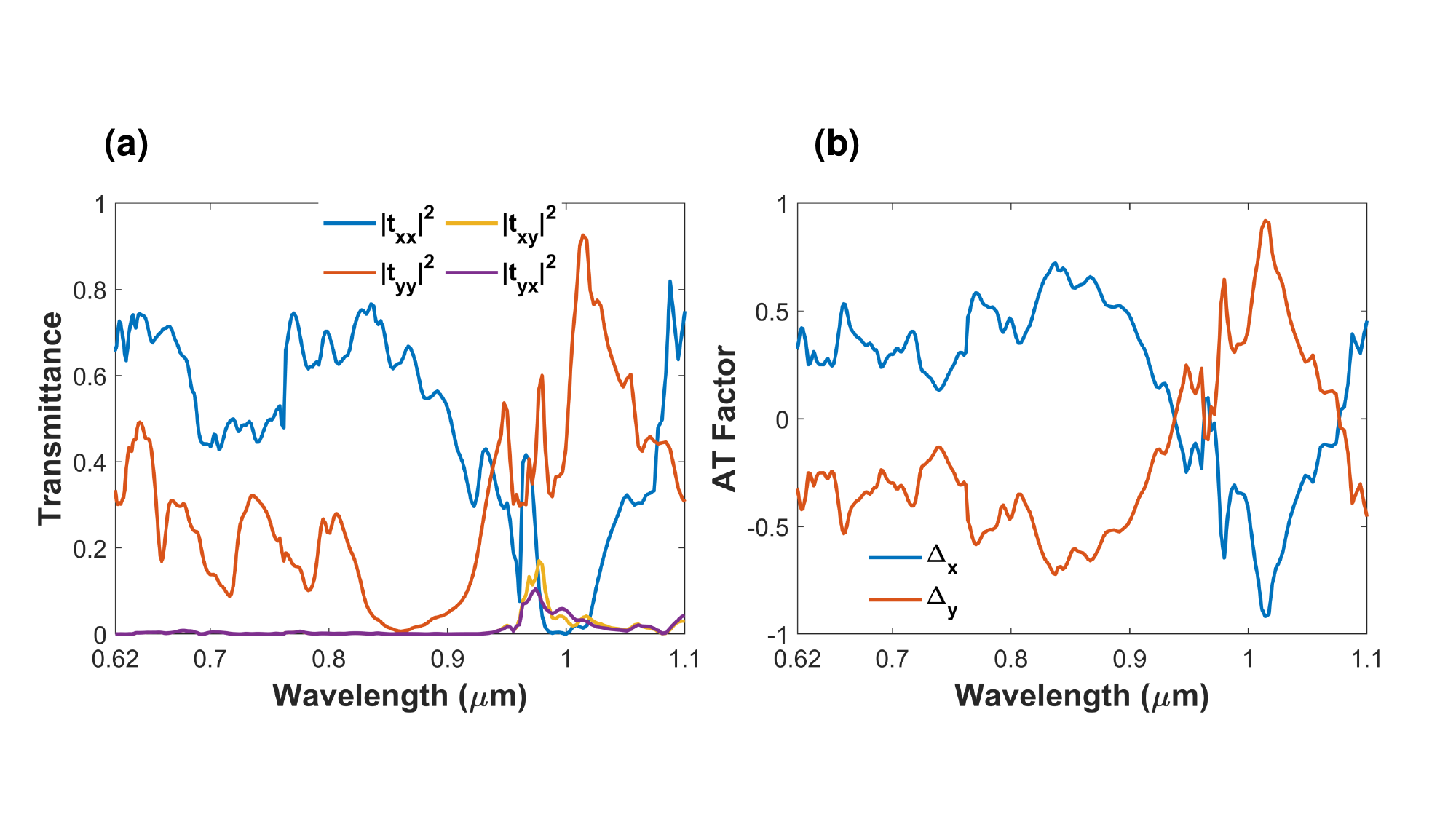}}
\caption{Spectral variation of the \textbf{(a)} magnitude squared transmission coefficients, and \textbf{(b)} AT factors for x ($\Delta_x$) and y ($\Delta_y$) polarized incident light.}
\label{fig:linear1}
\end{figure*}

Fig. \ref{fig:linear1}(a) illustrates the transmission spectra for co- and cross-polarization coefficients with a linearly polarized light incident along the z direction. The results show that $|t_{yy}|^2$ has a peak value of 0.930 at 1.014 $\mu$m wavelength, while $|t_{xx}|^2$ = 0.01 at this frequency. This feature of the spectral response indicated asymmetric transmission around this wavelength. We introduced AT factors for x ($\Delta_x$) and y ($\Delta_y$) polarized incident light to quantify the extent of AT. The quantities are defined as \cite{menzel2010asymmetric}
\begin{equation}
\label{eqn:AT}
    \Delta_x=|t_{xx}|^2+|t_{yx}|^2-|t_{yy}|^2-|t_{xy}|^2=-\Delta_y
\end{equation}

Fig. \ref{fig:linear1}(b) depicts the spectral variation of the two AT factors. As expected from Eq. \ref{eqn:AT}, $\Delta_x$ and $\Delta_y$ are equal and opposite. The results further show that $\Delta_y$ reaches a peak value of 0.92 at 1.01 $\mu$m. Hence, incident light with linear polarization experiences AT around this frequency.

\subsection{Linear polarization manipulation}

To quantify the cross-polarization conversion phenomena for x and y polarized incident light, we defined two polarization conversion ratios, $PCR_x$ and $PCR_y$ \cite{hao2007manipulating},
\begin{equation}
    PCR_x=\frac{|t_{yx}|^2}{|t_{yx}|^2+|t_{xx}|^2}  \hspace{10mm} PCR_y=\frac{|t_{xy}|^2}{|t_{xy}|^2+|t_{yy}|^2}, 
\end{equation}

\begin{figure*}
\centerline{\includegraphics[width=16cm]{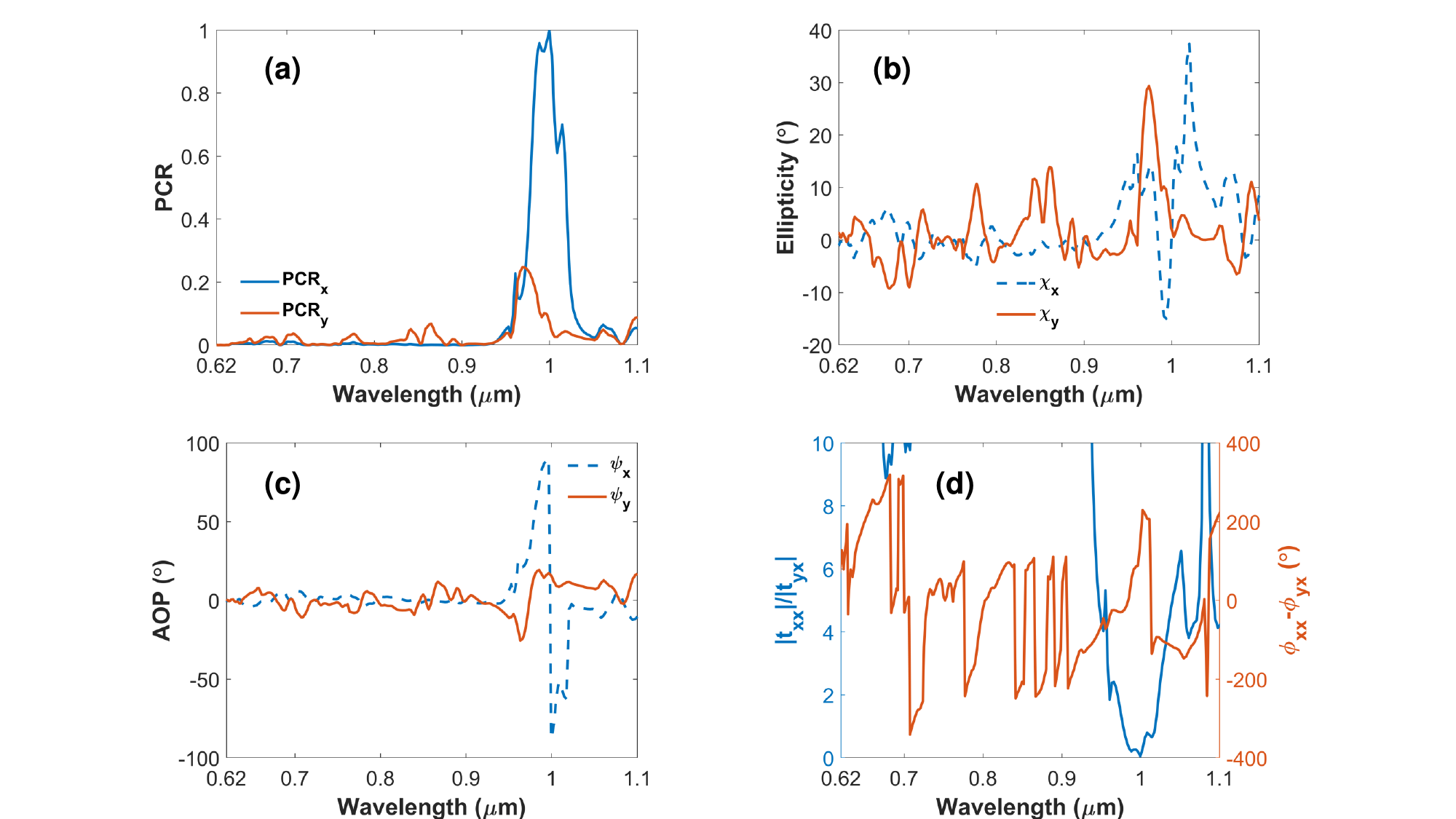}}
\caption{Spectral variation of \textbf{(a)} polarization conversion ratio (PCR), \textbf{(b)} ellipticity, and \textbf{(c)} AOP for x and y polarized incident light. \textbf{(d)} Relative values of the magnitude and phase of co- and cross-polarization transmission coefficients when the incident light is x polarized.}
\label{fig:linear2}
\end{figure*}

Fig. \ref{fig:linear2}(a) shows the $PCR$s as a function of incident light wavelength. $PCR_x$ has a peak of 0.997 at 1 $\mu$m wavelength. This means an x-polarized incident light almost perfectly gets converted to y-polarized light after transmission through the structure at a wavelength of 1 $\mu$m. $PCR_y$, on the other hand, has relatively tiny values throughout the considered spectral range. Hence, for y polarized light, there is no such cross-polarization conversion.  

We used the Stokes parameters to calculate the ellipticity $\chi$ and angle of polarization (AOP) $\psi$. The Stokes parameters ($S_{0,x(y)}$, $S_{1,x(y)}$, $S_{2,x(y)}$, and $S_{3,x(y)}$) were defined by the following set of equations for x(y) polarized light incidence \cite{goldstein2003polarized},

\begin{equation}
\label{eq:stokes}
\begin{aligned}
    & S_{0,x(y)}=|t_{xx(xy)}|^2+|t_{yx(yy)}|^2 ,\\ &
    S_{1,x(y)}=|t_{xx(xy)}|^2-|t_{yx(yy)}|^2 ,\\ & S_{2,x(y)}=2|t_{xx(xy)}||t_{yx(yy)}|\mathrm{cos}(\phi_{xx(xy)}-\phi_{yx(yy)}) \hspace{4mm}, \\ &
    S_{3,x(y)}=2|t_{xx(xy)}||t_{yx(yy)}|\mathrm{sin}(-\phi_{xx(xy)}+\phi_{yx(yy)}).
\end{aligned}
\end{equation}

Using the Stokes parameters defined in Eq. \ref{eq:stokes}, we calculated ellipticity ($\chi_{x(y)}$) and AOP ($\psi_{x(y)}$) based on the following set of equations for x(y) polarized illumination \cite{peng2018electromagnetic},

\begin{equation}
\label{eq:AOP}
    \mathrm{sin}(2\chi_{x(y)})=\frac{S_{3,x(y)}}{S_{0,x(y)}} \hspace{20mm}
    \mathrm{tan}(2\psi_{x(y)})=\frac{S_{2,x(y)}}{S_{1,x(y)}}
\end{equation}

Ellipticity and AOP are used to evaluate the extent of linear to elliptical (or circular) polarization conversion in the transmitted light. Both the quantities calculated from Eq. \ref{eq:AOP} are shown in Fig. \ref{fig:linear2}(b) and (c), respectively, for both x and y polarized light illumination as a function of wavelength. For x polarized incident light, ellipticity ($\chi_x$) has a peak value of 37.4$^\circ$ at 1.02 $\mu$m wavelength. At the same wavelength, AOP ($\psi_x$) is -5.65$^\circ$.  Ideally, a right circularly polarized (RCP) wave is expected to have an ellipticity of 45$^\circ$ \cite{khan2019linear}. This means the x-polarized incident wave gets converted to an RCP wave in the transmitted wave at a wavelength of 1.02 $\mu$m. This is further validated by calculating the relative values of the magnitude and phase of the co-polarization ($t_{xx}$) and cross-polarization ($t_{yx}$) transmission coefficients, as shown in Fig. \ref{fig:linear2}(d). The fact that at 1.02 $\mu$m wavelength, |$t_{xx}$|/|$t_{yx}$| = 1.3 and the phase difference, $\phi_{xx}$--$\phi_{yx}$ = --93.05$^\circ$, further indicates that the transmitted wave is RCP \cite{khan2019multiband}. Hence, our proposed structure can produce LP to CP conversion at 1.02 $\mu$m wavelength. In the y polarized incident light case, the ellipticity ($\chi_y$) had a peak value of 29.36$^\circ$ at 974.29 nm wavelength. This is a characteristic of elliptically polarized waves. Therefore, the y-polarized incident light with a wavelength of 974.29 nm was converted to an elliptically polarized wave upon transmission through the metasurface. 

\subsection{Circular dichroism and cross-polarization conversion effect}
We investigated the device's response under circularly polarized (CP) illumination. We considered the Jones matrix ($T_{circ}$) for CP incidence. The elements of the matrix can be related to the elements of ${T_{lin}}$ in Eq. \ref{eq:jones_lin} \cite{liu2015asymmetric},

\begin{equation}
\label{eq:jones_circ}
   T_{circ}=
    \left[ \begin{array}{cc}
        t_{rr} & t_{rl} \\
         t_{lr}& t_{ll}
    \end{array}
    \right]=\frac{1}{2}\left[ \begin{array}{cc}
        t_{xx}+t_{yy}+i(t_{xy}-t_{yx}) & t_{xx}-t_{yy}-i(t_{xy}+t_{yx}) \\
         t_{xx}-t_{yy}+i(t_{xy}+t_{yx})& t_{xx}+t_{yy}-i(t_{xy}-t_{yx})
    \end{array}
    \right].
\end{equation}

\begin{figure*}
\centerline{\includegraphics[width=1.1\textwidth]{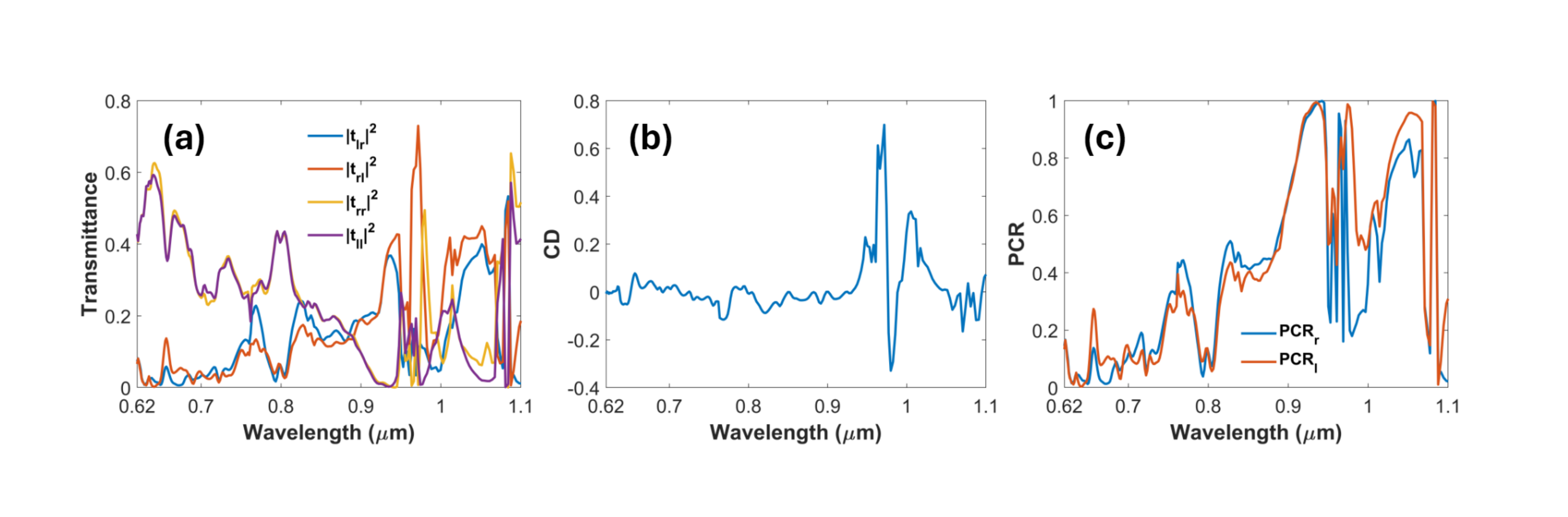}}
\caption{\textbf{(a)} Spectral variation of the magnitude squared transmission coefficients for RCP and LCP incidence, \textbf{(b)} CD spectra, \textbf{(c)} PCRs for circularly polarized incident waves.}
\label{fig:circ1}
\end{figure*}

In Eq. \ref{eq:jones_circ}, the subscripts \textit{r} and \textit{l} denote right (RCP) and left (LCP) circularly polarized light, respectively. The matrix entries $t_{i,j} (i,j=r,l)$ represent the transmission co-efficient for \textit{i} polarized light incidence when \textit{j} polarized component of the transmitted light is probed. Fig. \ref{fig:circ1}(a) depicts the variation of the matrix elements or the transmission coefficients over the considered spectral range. There is an asymmetry between the two cross-polarization coefficients ($|t_{rl}|^2$ and $|t_{lr}|^2$) around the 970 nm wavelength. To quantify this asymmetry, we introduced the parameter circular dichroism (CD), which is defined by,

\begin{equation}
\label{eq:CD}
    \mathrm{CD}=|t_{rl}|^2-|t_{lr}|^2+|t_{ll}|^2-|t_{rr}|^2.
\end{equation}

The spectral variation of the CD calculated from Eq. \ref{eq:CD} is shown in Fig. \ref{fig:circ1}(b). It peaked at a wavelength of 971.5 nm with a value of 0.7. Moreover, we investigated the ability of the star-shaped antenna-based structure to produce cross-polarization conversion for circularly polarized light. Similar to the case of linearly polarized light, we introduced PCR factors for RCP ($PCR_r$) and LCP ($PCR_l$) incidence as,

\begin{equation}
    PCR_r=\frac{|t_{lr}|^2}{|t_{lr}|^2+|t_{rr}|^2}  \hspace{10mm} PCR_l=\frac{|t_{rl}|^2}{|t_{rl}|^2+|t_{ll}|^2} 
\end{equation}

Fig. \ref{fig:circ1}(c) illustrates the two PCR factors. They had values close to 1 at two different bands: one around the 935 nm and another around the 1080 nm wavelength. This means the proposed metasurface can produce perfect cross-polarization conversion in transmitted light for both RCP and LCP at two different spectral bands.

\begin{table*}
\center
\caption{Comparison of our multifunctional chiral metasurface with previously reported relevant structures. \textdagger represents results from this work.}
\begin{tabular}{p{0.25\linewidth} p{0.13\linewidth} p{0.07\linewidth} p{0.07\linewidth} p{0.2\linewidth} p{0.09\linewidth}}

\hline 
\textbf{Structural layout} & \textbf{Operating wavelength range} & \textbf{CD} & \textbf{AT} & \textbf{Polarization conversion} & \textbf{Reference} 
\\
 
\hline

L shaped silicon antenna on $\mathrm{SiO_2}$/Gold & NIR & 0.75 & $\textrm{--}$ & $\textrm{--}$ & \cite{ali2023dielectric}
\\
Coupled gammadiation and Babinet structures
& NIR & 0.56 & 0.45 & $\textrm{--}$ & \cite{liu2019dual}

 \\

Au-$\mathrm{SiO_2}$-Au system with rectangular holes
& NIR & 0.76 & $\textrm{--}$ & $\textrm{--}$ & \cite{wu2022near}

\\

Two Al resonators embedded in $\mathrm{SiO_2}$ & Far-IR & 0.64 & 0.55 & Linear to elliptical and x to y & \cite{yin2023multi}

\\

Asymmetric Ag split-ring resonators embedded in dielectric  & Mid-IR & 0.1 & $\textrm{--}$ & Linear to elliptical & \cite{peng2018electromagnetic}

\\

Star-shaped Ag antennas embedded in Ag & NIR & 0.7 & 0.92 & (i) Linear to circular or elliptical, (ii) linear and circular cross-polarization conversion & \textdagger

\\

\hline
\quad
\end{tabular}
\label{table1}
\end{table*}

Table \ref{table1} compares the performance of the proposed star-shaped nanoantenna-based multifunctional chiral metasurface with previously reported relevant structures. Different performance parameters like CD and AT of our designed structure are comparable to previous works at the same spectral band. However, chiral platforms with diverse functionalities are limited to the longer wavelength portion of the electromagnetic spectrum. The NIR dual-functional structure of Ref \cite{liu2019dual} has a relatively lower CD and AT than other structures and does not incorporate the polarization conversion capability, which is key to availing several important chiral applications in optical communication systems. These comparisons further show the novelty of our proposed device as a multifunctional chiral platform incorporating a wide range of functionalities.  

\subsection{Electric field distributions}
\begin{figure*}
\centerline{\includegraphics[width=16cm]{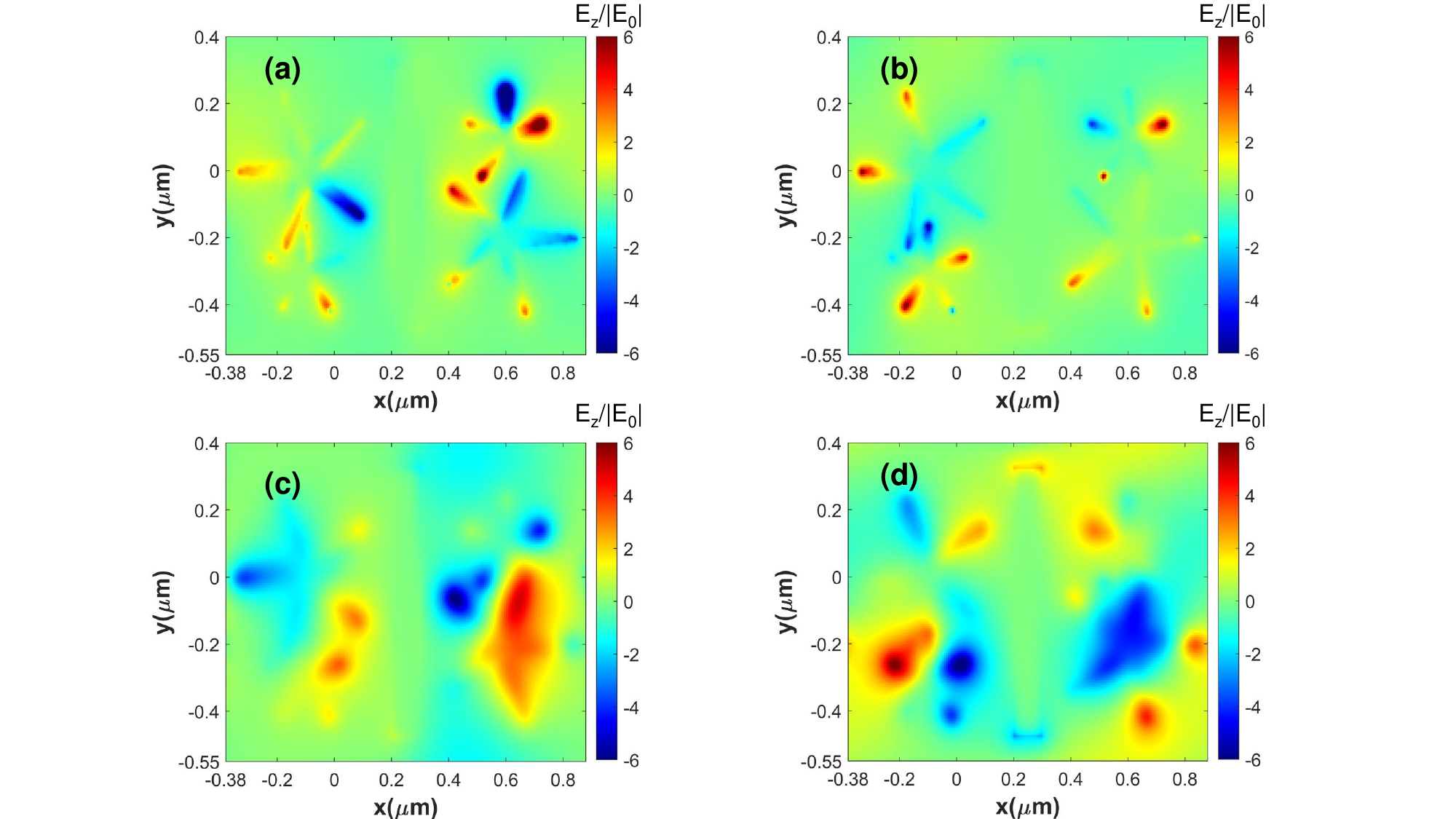}}
\caption{The x-y plane spatial distribution of z component of the electric field, $E_z$ for \textbf{(a)} x and \textbf{(b)} y polarized incident light with a wavelength of 1.01 $\mu$m, \textbf{(c)} LCP and \textbf{(d)} RCP incidence for a wavelength of 971.5 nm. All the fields have been normalized by the magnitude of the incident field. The x-y planes in each figure is positioned at the upper surface of the star-shaped antennas.}
\label{fig:field1}
\end{figure*}

To further investigate the origin of the handedness selective transmission asymmetry of the final structure, we simulated the electric field distributions of the structure at the relevant wavelengths. Fig.\,\ref{fig:field1}(a) and (b) illustrate the z-component of the electric field ($E_z$) distributions at 1.01 $\mu$m, where AT factor is maximum, for x and y polarized incident light, respectively. All the x-y plane cross-sections of the distributions are taken at the upper surface of the star-shaped antennas. The field distribution for the x-polarized light (Fig.\,\ref{fig:field1}(a)) shows a periodic alteration in sign along the y direction, indicating the presence of a surface plasmon mode. These oscillating charges along the y-direction produce a cross-polarized outgoing wave, quantified by $PCR_x$=1 around this frequency in Fig. \ref{fig:linear2}(a) \cite{peng2018electromagnetic}. A similar alteration in the sign of $E_z$ along the y direction can be observed in Fig.\,\ref{fig:field1}(b). This resulted in a $|t_{yy}|^2$ close to 1 at this frequency.   

The field distributions for LCP and RCP at a wavelength of 971.5 nm (where maximum CD appears) have been illustrated in Fig. \ref{fig:field1}(c) and (d), respectively. For RCP incident light, the distribution of $E_z$ shows an oscillating charge along the y direction, and the polarity alternates moving along the x direction. Moreover, charges of opposite polarity accumulate at the two ends of the metal bar, indicating the presence of a plasmon mode. In the case of the LCP polarized light, the oscillation of the $E_z$ polarity is not very prominent and does not change pattern along the x direction, as seen in Fig. \ref{fig:field1}(a). Since the RCP incident light couples more strongly to the plasmon modes, this will have a lower transmission than when LCP light is used as excitation, resulting in a CD of 0.7.

\subsection{Chiral sensing}
The determination of the handedness of chiral molecules is of paramount importance in applications such as drug production. One of the enantiomers is often pharmacologically significant while the other either has side effects or is functionless \cite{du2020chiral}. As a result, enantioselective drugs are usually preferred compared to their racemic mixture. Production of such single enantiomer drugs requires close investigation of molecular chirality at every step of the process through CD spectroscopy \cite{kelly2005study}. The inherently weak response from the standalone molecules makes the processing time-consuming \cite{kelly2005study} and calls for mechanisms to artificially enhance the CD spectra. Chiral metasurfaces have been extensively explored in past literature for ultrafast detection of molecule handedness by enhancing the interaction between light and chiral molecules \cite{kelly2018chiral,fan2021terahertz,hendry2010ultrasensitive}. 

To demonstrate the ability of the proposed structure in chiral sensing, we used 1,2-propanediol, a well-studied chiral molecule, as the test case. This molecule has two enantiomers, namely S-enantiomer and R-enantiomer \cite{bennett2001microbial}. The interaction between the chiral molecules and the metasurface was represented by a pair of dipoles oriented along the z-direction, consisting of one electric dipole with a magnitude of $3.54\times10^{-31}$ Cm and one magnetic dipole with a magnitude of $1.06\times10^{-22} \mathrm{Am^2}$, placed 10 nm away from the bottom surface of the metasurface. The scenarios with the two dipoles being in and out of phase with each other represent R and S enantiomers, respectively. A similar numerical setup was previously reported \cite{zhao2017chirality} for 1,2-propanediol with a good degree of agreement with experimental results in the NIR spectral band. 

\begin{figure*}
\centerline{\includegraphics[width=16cm]{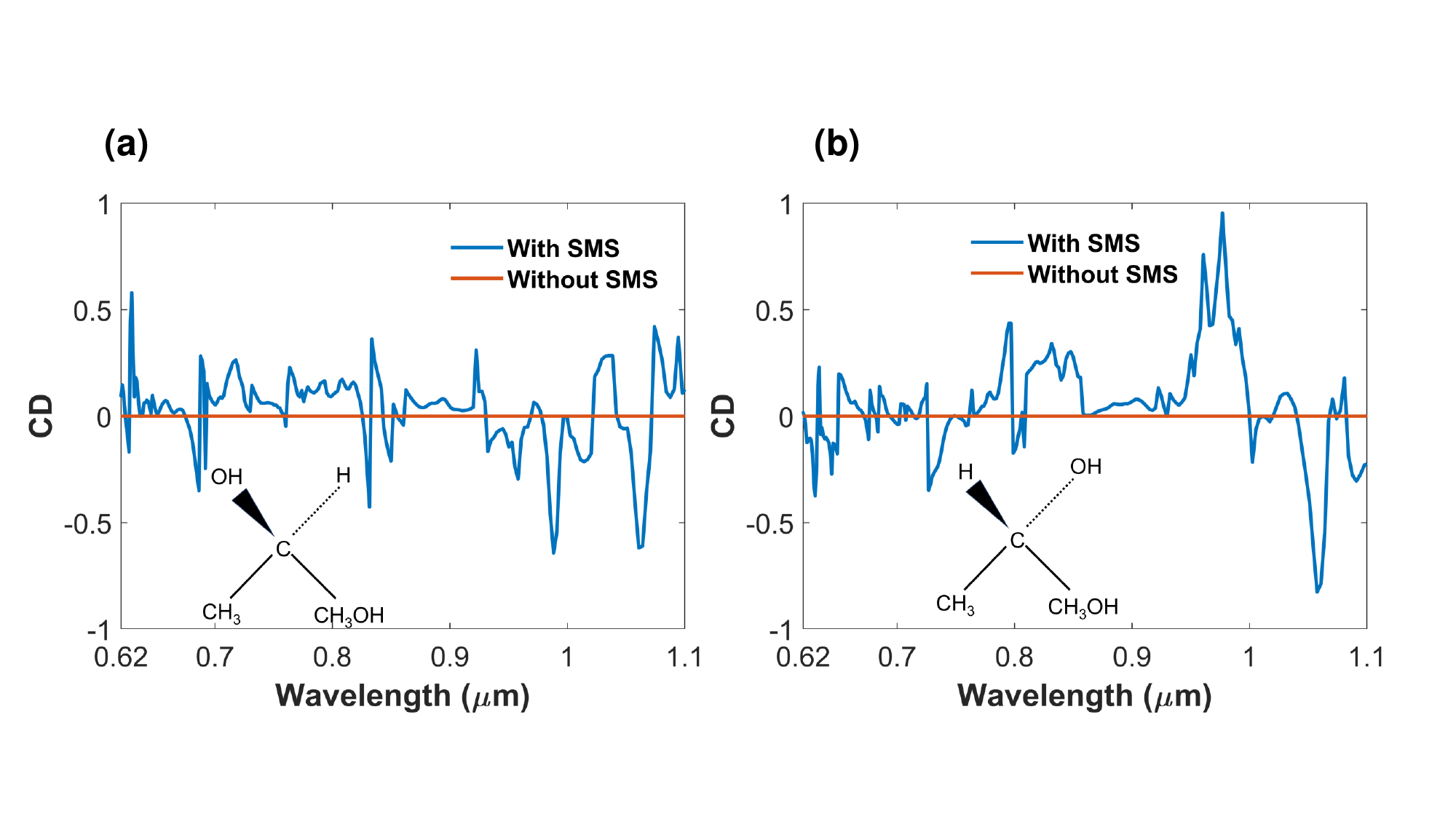}}
\caption{The CD spectra for \textbf{(a)} R and \textbf{(b)} S enantiomer of 1,2-propanediol with and without the interaction of the star-shaped antenna-based structure. The insets in (a) and (b) represent the chemical structures of the R and S isomers of 1,2-propanediol \cite{saxena2010microbial}.}
\label{fig:sensing}
\end{figure*}

The CD spectra for both enantiomers have been shown in Fig. \ref{fig:sensing} for the cases of a standalone chiral molecule layer and a chiral molecule layer interacting with the proposed chiral metasurface. In the absence of the metasurface, the CD signal is negligible. However, when the molecule is placed close to the metasurface, we achieved a CD peak of -0.64 at 988 nm wavelength for the R-enantiomer (Fig. \ref{fig:sensing}(a)) and 0.98 at 954 nm wavelength for its S counterpart (Fig. \ref{fig:sensing}(b)). These results show the excellent ability of the proposed metasurface to enhance the CD signal of the individual metamolecule and to differentiate the enantiomers optically. This ability to distinguish enantiomers of a chiral molecule can be crucial in enantiospecific manufacturing applications. For instance, the R-isomer of 1,2-propanediol is used to prepare a drug called Levofloxacin \cite{nagaoka2019drug}. This drug is extensively used as a broad-spectrum antibiotic for treating humans across the globe \cite{croom2003levofloxacin}. The other enantiomer of the drug is known to produce neurotoxicity in the patient \cite{coelho2021enantioselectivity}. Hence, it is essential to differentiate the R and S enantiomers of one of its precursors, 1,2-propanediol. Therefore, our designed chiral metasurface can find applications in the production process of Levofloxacin and enantiospecific drugs.

\section{Conclusion}
We proposed a star-shaped Ag antenna-based multifunctional chiral metasurface for the NIR regime. The final architecture was chosen through extensive optimization and comparison of three different topologies of the nanoantennas. This final structure showcased handedness selective asymmetry in transmission for both linear and circular polarizations, with maximum values of 0.7 at 971.5 nm and 0.92 at 1.01 $\mu$m wavelength, respectively. These values aligned with previously reported state-of-the-art chiral meta-devices for the NIR spectral band. The investigation of the electric field profiles revealed the presence of plasmons as the underlying cause of such asymmetry toward polarization handedness. Moreover, the multifunctional structure manifested several polarization manipulation functionalities in the FDTD simulation environment, namely cross-polarization conversion for both linear and circular polarizations and linear to circular and elliptical polarization conversion at the NIR frequencies. Chiral metasurfaces with such a diversified range of functionalities have not been reported previously for the NIR range. Furthermore, the ability of the designed chiral structure to differentiate two enantiomers of 1,2-propanediol suggested its potential in the pharmaceutical and biomolecular industry as a tool for chiral sensing. The successful integration of several NIR chiral functionalities in the same structure reported here is expected to pave the way for compact photonic systems for diverse NIR sensing, imaging, communication, and biochemistry applications.

\begin{acknowledgements}

M. E. Karim and A. Zubair thank the Department of Electrical and Electronic Engineering (EEE) at Bangladesh University of Engineering and Technology (BUET) for support and facilities.

\end{acknowledgements}

\section*{Disclosures}
The authors declare that they have no known competing financial interests or personal relationships that could have appeared to influence the work reported in this paper.

\section*{Authorship Contributions}
\textbf{Md. Ehsanul Karim}: Conceptualization, Formal analysis, Methodology, Visualization, Software, Investigation, Writing - original draft.\\
\textbf{Ahmed Zubair}: Supervision, Conceptualization, Methodology, Project administration, Resources, Visualization, Writing - original draft, Writing-review \& editing.

\section*{Data Availability Statement}

The data that support the findings of
this study are available from the
corresponding author upon reasonable
request.

\bibliography{Main}

\end{document}